\documentstyle[twoside,fleqn,psfig]{an-art}
\begin{document}
\def\address #1END {{\vspace{9mm}\noindent\small Address of the author: \medskip \\ #1}}
\def\addresses #1END {{\vspace{9mm}\noindent\small Addresses of the authors: \medskip \\ #1}}
  \def\bruch#1#2{{#1 \over #2}}
  \def\Mpunkt{\dot{M}}
  \def\Msun{M_{\odot}}
  \def\Lsun{L_{\odot}}
  \def\pism{p_{\scriptscriptstyle I\!S\!M}}
  \def\aquer{\scriptstyle{\langle} a \scriptstyle{\rangle}}
  \def\nH{n_{\scriptstyle{\langle} H \scriptstyle{\rangle}}}
  \def\abl#1#2{\bruch {d #1}{d #2}}
  \def\pabl#1#2{\bruch {\partial #1}{\partial #2}}
\def\ga{\mathrel{\mathchoice {\vcenter{\offinterlineskip\halign{\hfil
$\displaystyle##$\hfil\cr>\cr\sim\cr}}}
{\vcenter{\offinterlineskip\halign{\hfil$\textstyle##$\hfil\cr
>\cr\sim\cr}}}
{\vcenter{\offinterlineskip\halign{\hfil$\scriptstyle##$\hfil\cr
>\cr\sim\cr}}}
{\vcenter{\offinterlineskip\halign{\hfil$\scriptscriptstyle##$\hfil\cr
>\cr\sim\cr}}}}}
\begin{titlepage}
\setcounter{page}{1}
\def\makeheadline{\vbox to 0pt{\vskip -30pt\hbox to 50mm
{\small Astron. Nachr. 318 (1997) 1, 1--x \hfill}}}
\makeheadline
\title {Cyclotron spectroscopy of VV Puppis }
\author{{\sc  Axel D.~Schwope}, Potsdam-Babelsberg, Germany \\
\medskip
{\small Astrophysikalisches Institut Potsdam} \\
\bigskip
{\sc   Klaus Beuermann}, G\"ottingen, Garching, Germany \\
\medskip
{\small Universit\"{a}tssternwarte G\"{o}ttingen\\[1ex]MPI f\"{u}r 
extraterrestrische Physik}
}
\date{Received 1996 November 22; accepted } 
\maketitle
%
\summary
We present phase-resolved spectrophotometric observations 
of VV Puppis obtained during two different states of accretion. We 
confirm the detection of cyclotron lines from emission regions at both 
poles (Wickramasinghe et al. 1989) having significantly different field 
strength of $B_1 \simeq 31$~MG and $B_2 \simeq 54$~MG. Our phase-resolved 
data allowed the detection of phase-dependent wavelength shifts of the 
cyclotron lines from the main accretion pole which is 
due to the varying aspect of the observer. A corresponding motion of the 
cyclotron lines from the secondary pole appears likely. \\
Compared to 1984, 
the cyclotron lines from the main pole appeared redshifted in 1989, 
during an episode of higher system brightness. This shift 
can be explained assuming either different locations of the accretion 
spot and, hence, variations of the magnetic field, or variations of the 
plasma temperature both triggered by variations of the mass accretion 
rate.\\
The cyclotron lines from the second pole do not show such changes 
suggesting that the coupling region for the weakly accreting pole 
is more or less stationary whereas the coupling
region of the main pole varies in space, most likely
depending on the mass accretion rate.END
\keyw
cataclysmic variables --- AM Herculis binaries --- stars: 
individual (VV Pup) --- cyclotron radiation --- stars: 
magnetic fieldEND
\AAAcla
119; 120; 122END
\end{titlepage}

\kap{Introduction}
VV Puppis was recognized as third AM Herculis binary by the detection 
of strong and periodic variable linear and circular polarization by Tapia 
(1977). 
A high orbital inclination of $i \sim 75^\circ$ and a high ``southern'' 
colatitude of the main accretion region of $\delta \sim 150^\circ$ lead 
to a long selfeclipse of this main spot by the white dwarf itself as the 
system rotates. Visvanathan and Wickramasinghe (1979) and Stockman et 
al. (1979, hereafter SLB) detected intense low-frequency modulations of 
the continuum spectra during the bright phase, with the minima originally 
explained as cyclotron absorption troughs in a field of $\sim 30$~MG. 
Later, when detailed cyclotron models became available, the maxima in 
intensity were interpreted as cyclotron line emission and the field 
strength was redetermined to 31.5 -- 32.0 MG (Wickramasinghe \& 
Meggitt 1982; Barrett \& Chanmugam 1985).

Photometric variations and nonvanishing circular polarization during the 
faint phase, when the main accreting pole is out of view, have been 
interpreted by Liebert and Stockman (1979) as due to accretion in the 
vicinity of the second pole. Wickramasinghe et al. (1989, hereafter WFB) 
confirmed this supposition by the detection of two independent systems 
of cyclotron lines originating from both accretion regions. The field 
strength in the secondary region was found to be surprisingly high, 
$B_2 \simeq 56$~MG, and the newly determined field strength for the 
primary pole, $B_1 = 30.5$~MG, was lower than during the initial 
observations. These findings have been interpreted assuming that 
accretion occurs at the footpoint of the same closed field line in 
a decentered dipole field ($d_{\rm off} \simeq 0.1 R_{\rm wd}$). The 
lower value of $B_1$ derived from the more recent observations were
explained by WFB as consequence 
of a migration by $\sim 10^\circ$ of the main accretion 
region towards the magnetic equator.

\begin{figure}[t]
\begin{center}
\begin{minipage}{120mm}
\psfig{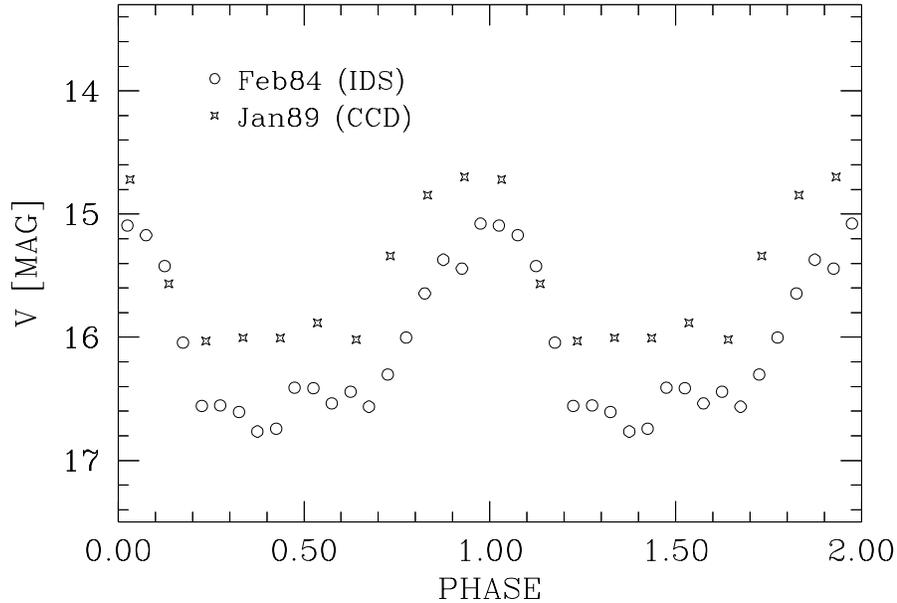}
\end{minipage}
\end{center}
\caption{\label{lc}
Approximate V-band lightcurves for VV Pup as derived from 
spectrophotometric observations obtained on February 5, 1984, and on 
January 19, 1989. The same data are plotted over two orbital cycles. 
The phase convention is that given by Walker (1965).}
\end{figure}

In this paper, we describe the results of phase-resolved low-resolution 
spectrophotometry (cyclotron spectroscopy) of VV Pup obtained at two 
different epochs. The system was observed in different states of 
accretion and in both cases displayed cyclotron lines from both 
accretion regions. Cyclotron spectroscopy of AM Herculis cataclysmic 
binaries accompanied by simple (isothermal) cyclotron model calculations 
has proven its capability as an efficient plasma-diagnostic tool in 
order to derive the main plasma parameters, the basic accretion geometry 
and the field structure in e.g. UZ For and V834 Cen, respectively 
(Schwope et al. 1989; Schwope \& Beuermann 1990). For the case of VV Pup 
we analyse in this paper the motion of individual cyclotron lines and 
show that accretion in the two regions probably does not occur at 
the footpoints of the same closed 
field lines. This implies that the 
accretion regions at both hemispheres are probably 
fed by streams starting from different locations in the orbital plane.
As in the previous analysis of WFB, we found the cyclotron lines from the 
main accretion region wavelength-shifted between our different observations 
indicating either magnetic field or temperature variations. 

In chapter 2, we describe our spectrophotometric observations. The first 
section of chapter 3 deals with the photometric and the overall spectral 
behaviour. The motion of the cyclotron lines is analyzed in section 3.2,  
while a detailed analysis of the faint and bright phase cyclotron spectra 
is given in sections 3.3 and 3.4, respectively. Preliminary results of 
this work were presented in Schwope (1990).

\kap{Observations}
VV Pup was observed by us spectroscopically on February 5, 1984, 
using the ESO 3.6m telescope as part of an international campaign 
for multi-wavelength monitoring simultaneous with EXOSAT. The X-ray 
observations were further accompanied by optical photometric observations 
in the Walraven system using the ESO 90cm telescope and by optical 
polarimetry from the Sutherland site of the SAAO. Results of the X-ray 
observations were presented by Osborne et al. (1984) while the 
polarimetric observations were analyzed and published by Cropper \& Warner 
(1986).

The February 5, 1984, spectroscopic observations lasted from 0:40 to 3:30 
UT and were performed under photometric conditions. The Boller \& Chivens 
spectrograph with Image Dissector Scanner (IDS) was used. The wavelength 
range covered was 3850-8350~\AA\ with a FWHM resolution of $\sim 12\,$\AA. 
A total of 248 spectra were taken with integration times of $30\,$s or 
$60\,$s, interrupted by regular calibration exposures using a He-Ar 
lamp. The results of the simultaneous X-ray and optical spectroscopic 
observations will be published elsewhere. For the purpose of the pesent 
paper, we make use only of the shape of the continuum after collecting 
the spectra into 20 phase bins. The phase convention used throughout 
this paper is that given by Walker (1965) were phase zero refers to 
the maximum of the optical light curve. 

\begin{figure}[t]
\begin{center}
\begin{minipage}{120mm}
\psfig{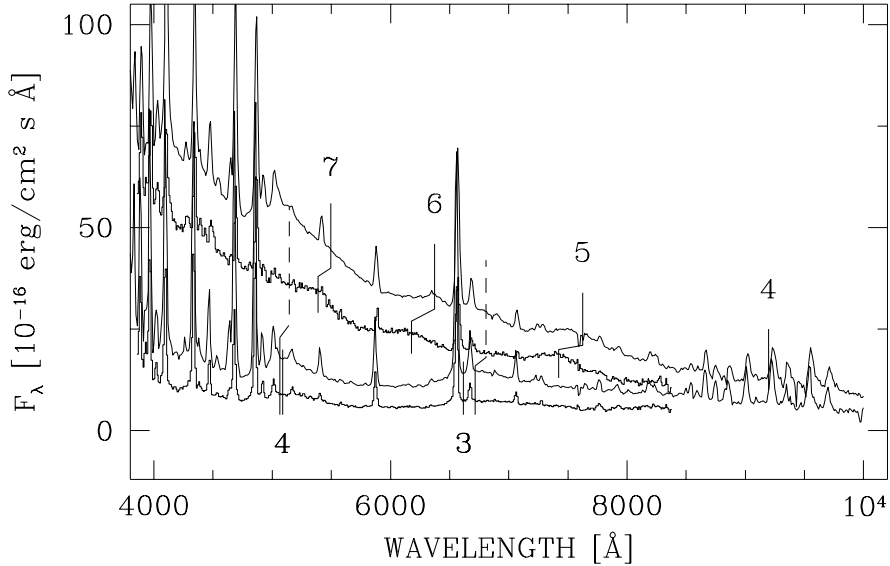}
\end{minipage}
\end{center}
\caption{\label{mean_spec}
Representative spectra for the centers of the bright and 
faint phases from both observational campaigns. The CCD-data extended 
farther into the red than the IDS-data. Final positions and shifts 
of individual cyclotron harmonics are given. Numbers below the spectra 
indicate harmonics from the second, weakly accreting pole.}
\end{figure}

VV Pup was reobserved spectroscopically on January 19, 1989, using 
the ESO/MPI 2.2 m telescope at La Silla, Chile. A total of 10 spectra 
with integration times of 575 sec were gathered between 5:47 and 
7:29 UT under photometric conditions with seeing of approximately 2~arcsec. 
The Boller \& Chivens spectrograph was equipped with ESO grating 
13 with 150 grooves/mm yielding a reciprocal dispersion of 450~\AA/mm. 
The spectra were recorded using a RCA-CCD with pixel 
size of 15~$\mu$m ($15.6 \times 9.8\,{\rm mm}^2$). The CCD images were 
binned on chip $2 \times 2$ before readout. Using a 2'' slit, we 
achieved a spectral resolution of $\sim 20$~\AA\ FWHM and covered 
the spectral range between 3500 - 10000~\AA.

All spectra described here were flux-calibrated using spectrophotometric 
standard stars. Since the projected slit width of the spectrographs 
matched with the seeing disk of the star, the spectrophotometric 
accuracy achieved is of the order of 10-20~\%. This could be directly 
checked for our 1984 observations by comparison with the simultaneously 
recorded Walraven photometry. 
Henceforth we refer to the 1984 spectrophotometric data as 'IDS-data' 
and to the 1989 data as 'CCD-data'. 

\kap{Results and analysis}
\sect{Optical brightness variations and overall spectral behaviour}
By folding individual spectra with the transmission curve for a standard 
Johnson V-filter, the light curves shown in Fig.~\ref{lc} were generated. The 
overall system brightness was significantly different on both occasions 
while the length of the bright phase $\Delta\phi_{\rm B}$ was quite 
similar. For the IDS-data and the CCD-data, we obtain $V_{\rm orb} = 16\fm8 
- 15 \fm 0$, $\Delta\phi_{\rm B} = 0.50 \pm 0.03$, and $V_{\rm orb} = 
16 \fm 0 - 14\fm7$, $\Delta\phi_{\rm B} = 0.49 \pm 0.05$, respectively. 
For comparison, the V-magnitude during a low state without significant 
accretion was $V \simeq 17\fm5$ (Bailey 1978; Liebert et al. 
1978). The light curves are slightly asymmetric with a slower increase 
than decrease as is typical for this system. Photometric maximum 
occurs around phase zero while the bright phase is centered around 
$\phi = 0.93$. The bright phase ends just before $\phi = 0.2$ which 
seems to be a general photometric feature of VV Pup independent of 
brightness (Cropper \& Warner 1986).

Representative spectra for the centers of the bright and faint phases, 
$\phi = 0.95$ and 0.45, are shown in Fig.~\ref{mean_spec}. 
Cyclotron lines from two 
accretion regions are clearly resolved. The final identification of 
individual harmonics from both poles are given in the figure. The lines 
originating at the weakly accreting second pole are discernible 
throughout the orbital cycle. The bright-phase continuum of the 
CCD-observations is significantly influenced and modulated by 
radiation from the second pole. The 
bright-phase cyclotron lines during the IDS-observations, when the 
system brightness was reduced, shine up more clearly and may be more 
easily distinguished from the faint cyclotron lines from the second 
pole. 

\begin{figure}[t]
\begin{center}
\begin{minipage}{120mm}
\psfig{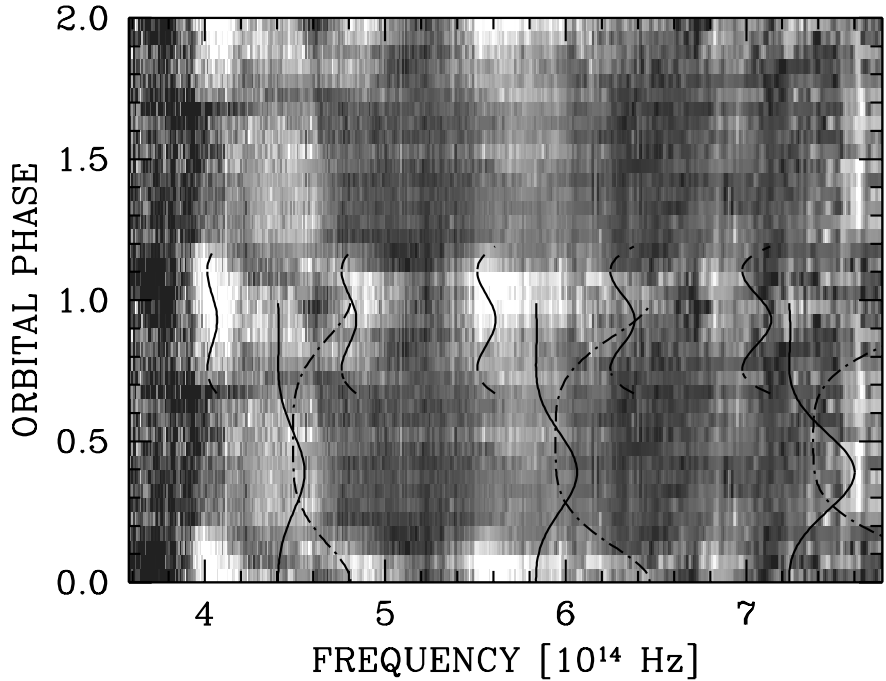}
\end{minipage}
\end{center}
\caption{\label{grey_spec}
Grey-scale representation of the phase-folded, continuum-subtracted
IDS-data. The phase resolution is 0.05 and all data are shown 
twice for clarity. Cyclotron harmonic emission is detected
at bright (around phase 1) and faint phases and appears white. The humps
are identified with the 5$^{\rm th}$ to 8$^{\rm th}$ harmonic in the bright
phase, and with harmonics 3 and 4 in the faint phase. Solid
lines mark expected positions of cyclotron maxima as a function of phase
for the magnetic fields and projection angles as given in the text for our
model 1. Results of our model 2 are shown using dashed-dotted lines. For the 
low-field primary pole these lines have been extended over the 
phase interval of nominal visibility (polar angle $90^\circ$, maximum 
redshift of the cyclotron lines) down to polar angles of $105^\circ$ in order
to demonstrate the effect of extended visibility due to inclined field lines 
as deduced from polarimetry.
}
\end{figure}

\begin{figure}[t]
\begin{center}
\begin{minipage}{120mm}
\psfig{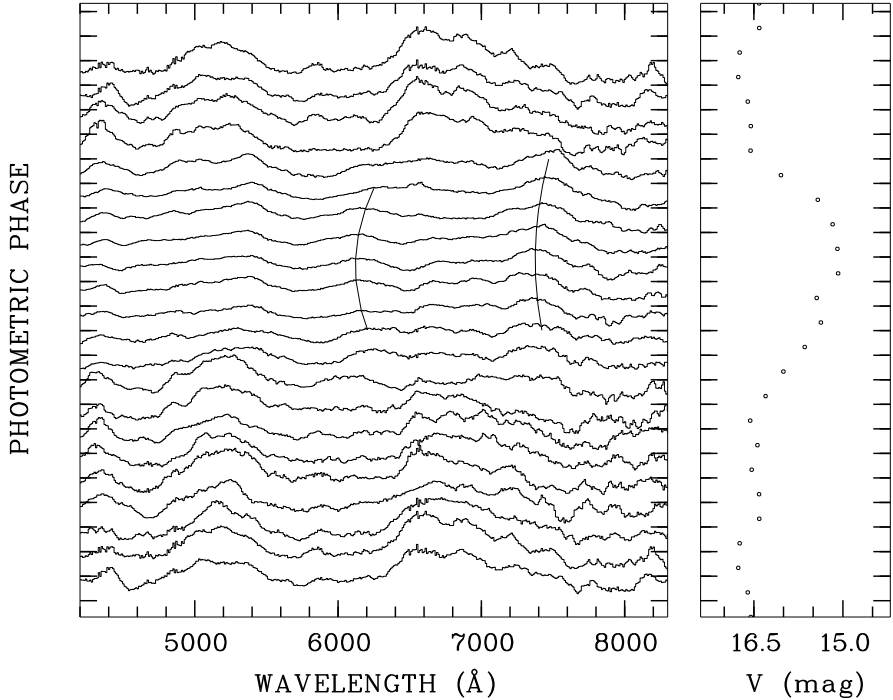}
\end{minipage}
\end{center}
\caption{\label{phas_spec}
Phase-resolved normalized spectra of VV Pup obtained on 
February 5, 1984 (IDS-data). The spectra were normalized by division of the 
original spectra by smooth curves approximating the global continuum.
The phase resolution is 0.05. In order to 
facilitate distinction between bright- and faint-phase spectra, the 
corresponding brightness information is given in the right-hand panel. 
Individual spectra were plotted with a vertical offset of 0.3 flux units
(relative units, continuum intensity = 1). Lines 
connecting individual humps are drawn for the 5$^{\rm th}$ and 6$^{\rm th}$ 
harmonics.}
\end{figure}

The maxima of the bright-phase cyclotron lines appear to be displaced 
between our two observations. This is not evident at the present stage 
of analysis (Fig.~\ref{mean_spec}) but becomes clear after subtraction of the 
faint-phase contributions from the bright-phase flux. This is not a 
straightforward procedure and the further analysis of these features is 
delayed to chapter 3.4. While the bright-phase cyclotron lines are 
displaced between the IDS- and CCD-observations, the faint-phase lines 
appear at the same position at given phase. 

The detection of cyclotron lines from both poles was first reported 
by WFB during an episode of brightness comparable to that during our 
CCD-observations. In fact, the positions of cyclotron maxima, both 
during the bright and faint phases, agree between the two sets of data. 
WFB could not detect any motion of harmonic peaks with phase. However, 
encouraged by or analysis' of the AM Herculis binaries UZ For,
V834 Cen, MR Ser, QS Tel and RX\,J0453--4213 
(Schwope \& Beuermann 1990, Schwope et al.~1990, 1993, 1995, Burwitz et 
al.~1996), which all show moving
cyclotron lines as predicted by cyclotron theory, 
we decided to take a closer look on our phase-resolved data. For that 
purpose, atomic emission lines were approximated by single or double 
Gaussians and subtracted. Each spectrum was then divided by a low-order 
polynomial fit to its continuum (typically of order 3). The result of 
this procedure for the phase-resolved IDS-data is shown in 
Figs.~\ref{grey_spec} (continuum-subtracted) and \ref{phas_spec} 
(continuum-divided) with a resolution of 0.05 phase units. 
In the diagram showing the continuum-divided spectra, the intensity modulation 
during the faint phase appears enhanced with respect to the bright phase
due to the smaller numbers used for division. From both Figures, 
\ref{grey_spec} and \ref{phas_spec}, 
it becomes clear that the cyclotron line system of the main accretion 
pole moves with phase with maximum blueshift roughly in the center 
of the bright phase (best visible in the 5$^{\rm th}$ and 6$^{\rm th}$ 
harmonics). Similarly the lines from the secondary pole are seemingly
moving with maximum blueshift around phase 0.40 but this identification 
causes some problems with the probable location of this region on the 
white dwarf (see next Sect.) and we regard this identification as tentative
only.

In order to isolate the cyclotron spectrum of 
selfeclipsing accretion regions, the simplest procedure is
to subtract a mean faint-phase spectrum from the bright-phase spectra.
In case of VV Pup, however, this procedure would lead to erroneous 
results because the cyclotron lines from the second pole are shifted 
during the bright phase. We, therefore, subtracted an appropriate 
synthesized cyclotron spectrum from the bright-phase data. The details are 
described in Sect.~3.4.

Finally, we note that the phase-dependent motion of both cyclotron 
line systems may be recognized clearly only in the IDS-data. The CCD-data 
seem to display the same motion of the lines from the second pole but the 
coarse phase resolution and the relative weakness (in terms of modulation 
depths of individual harmonics, not in terms of absolute flux) 
of the lines from 
the main pole prevent a phase-resolved study during the bright phase. 

\sect{The motion of cyclotron lines}
In a weakly relativistic plasma, maximum intensity of individual 
harmonics for a given field strength depend on temperature and polar 
angle $\theta$, i.e. the angle between the magnetic field and the observer. 
The observed motion of cyclotron lines during an orbital revolution is, 
therefore, caused by the varying aspect of the accreting field line(s) 
as the white dwarf rotates. In Schwope \& Beuermann (1990), we have 
described how $\theta$-dependent cyclotron models may be parameterized 
and applied to the observed motion of cyclotron peaks. Free fit parameters 
then are the phase of maximum blue- or redshift $\phi_0$, the orbital 
inclination $i$, the inclination of field line(s) $\eta$ fixed in the 
frame of the white dwarf, the plasma temperature kT, and the magnetic 
field strength $B$. The angles $i$ and $\eta$ combined with the orbital 
phase result in an expression for the polar angle $\theta$. In the 
following subscripts '1' and '2' are added to quantities belonging to
either the main (1) or the secondary (2) accreting pole.

The positions of individual harmonics observed in VV Pup have been 
determined interactively on a graphic display via cursor input. Since 
we have only relatively few data points, the information on $i, \eta$, 
kT and $B$ which may be obtained by fitting our parameterized cyclotron 
models to the observed motion is not superior to that which is already 
known from polarimetry (applies to $i$ and $\eta$) or which may be 
obtained from fits to the flux distribution of individual spectra 
(applies to kT and $B$, see the next sections). The main interest in 
studying the motion of the lines then arises from the opportunity to 
fix the azimuth of the accreting field lines in both accretion regions. 
The phase of maximum blueshift 
of the 5$^{\rm th}$ 
and 6$^{\rm th}$ harmonics from the main pole coincides with the center 
of the bright phase, $\phi_{\rm C1} = 0.93 \pm 0.01$. 
Maximum blueshift of the 3$^{\rm rd}$ harmonic from the 
second pole seems to occur at  $\phi_{\rm C2} = 0.39 \pm 0.01$, where we have
identified the enhanced emission between harmonics 5,6 and 7,8 of the
main pole as redshifted $3^{\rm rd}$ and $4^{\rm th}$ harmonics from the
secondary pole. 
These numbers imply that the line systems move 
nearly in antiphase. The phase difference $\phi_{\rm C2} 
- \phi_{\rm C1}$ corresponds 
to an azimuthal difference of $\Delta\psi_{12} = 165^\circ \pm 10^\circ$ 
with the primary pole leading. In Fig.~\ref{grey_spec} we show (as solid lines,
referred to as model 1) 
the predicted positions of cyclotron maxima of the two
accretion regions where we have assumed an orbital inclination $i = 75^\circ$,
magnetic colatitudes for the main and the secondary regions of $\eta_1
= 150^\circ$, $\eta_2 = 20^\circ$, plasma temperatures in equivalence to
$kT_1 = 10$\,keV, $kT_2 = 5$\,keV and field strengths $B_1 = 31$\,MG and
$B_2 = 54$\,MG, respectively. These models show that 
cyclotron harmonics cannot reliably be identified in the blue spectral 
range ($\nu 
\geq
6.6 \times 10^{14}$\,Hz) because of residual, unresolved
atomic emission lines.

The azimuths of field lines $\psi_{\rm F1}, \psi_{\rm F2}$ in both 
emission regions (which may deviate from the azimuths of the corresponding 
accretion spots $\psi_{\rm S1}, \psi_{\rm S2}$) can be located in the 
binary reference frame, using the phase of inferior conjunction of the 
secondary star as determined by Cropper (1988), $\phi_{\rm MS} = 0.067 
\pm 0.030$. From $\psi_{\rm F1,F2} = \phi_{\rm MS} - \phi_{C1,C2}$, 
one obtains $\psi_{\rm F1} \simeq 50^\circ$ and $\psi_{\rm F2} \simeq 
245^\circ$ where the value for $\psi_{\rm F1}$ agrees with that of Cropper 
(1988). He derived $\psi_{\rm F1}$ from the phase difference $\phi_{\rm MS} - 
\phi_{\rm CP}$ between inferior conjunction and 
the center of the circularly positive polarized phase interval. The agreement
between polarimetric and spectroscopic results concerning $i$, $\eta$ and 
$\phi_{\rm C1}$ is excellent, giving further confidence in the applicability
of simple cyclotron models for this purpose. This is particularly important
since Piirola et al.~(1990) found evidence for a linearly extended 
emission region at the main pole. This could distort the profiles of
individual harmonics and make motions difficult to interpret. Obviously 
this applies not strongly to the IDS-data but it might be important
for the CCD-data which were obtained when the system was brighter. Perhaps 
this is one of the reasons (besides the lower time resolution) why we 
could not detect motions of the cyclotron lines in this more recent
data set.

The center of the X-ray bright phase determines the azi\-muth of the 
primary accretion spot, $\psi_{\rm S1} \simeq 35^\circ$. The azimuthal 
angles of the spot and the field, 
$\psi_{\rm S1}$ and $\psi_{\rm F1}$, thus differ by $\sim 
15^\circ$, which simply means, that the field direction 
in the spot is not perpendicular to the surface (as expected). 
We have no observational quantity which may be used to 
determine the azimuth of the secondary spot $\psi_{\rm S2}$. If we then 
assume for simplicity that 
$\psi_{\rm S2} \simeq \psi_{\rm F2} \simeq 245^\circ$, a peculiar location 
of the secondary region is implied, nearly $3/4$ around the white dwarf. 
Such a geometry seems to be very unlikely, and it is presently not clear where 
the coupling region between magnetic field and the secondary stream 
should be located. We have two possible explanations: either the field
structure deviates significantly from a dipole so that the orientation of 
the local field in the spot gives no direct hint to the global orientation
of the field and to the likely azimuth of the spot or our
identification of the motion of the cyclotron lines is in error. With only
two measurements of the local field (strength and likely direction) 
in the two accretion regions we 
cannot explore the former explanation, in order to illustrate one possible 
solution within the latter we show in Fig.~\ref{grey_spec} with dashed-dotted
lines the predicted motion of a cyclotron line system in a field of 
$B_2 = 55$\,MG using $\phi_{\rm C2} = 0.0$ (referred to as model 2). 
This model implies that 
during the bright phase both cyclotron line systems are strongly 
blended and thus leaves the enhanced emission between harmonics 5,6 and 7,8
unexplained. Thus both possible solutions have their pro's and con's 
and the puzzle cannot be resolved here due to the lack of relevant 
data (e.g.~high S/N polarimetry or spectropolarimetry during the faint phase).
We assume in the following that the motion of the secondary cyclotron lines
can be described by model 1 (solid lines in Fig.~\ref{grey_spec}).


\begin{figure}
\begin{center}
\begin{minipage}{120mm}
\psfig{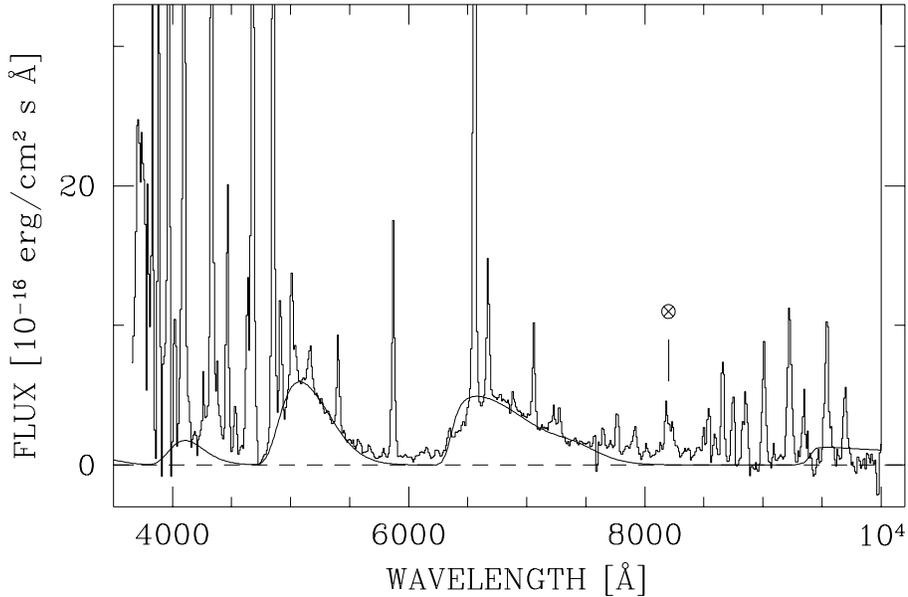}
\end{minipage}
\end{center}
\caption{\label{cyc_fai}
Cyclotron spectrum of the faint phase, $\phi = 0.45$, derived 
from CCD-data by subtraction of a smooth continuum. Also shown is an 
isothermal cyclotron model for kT = 5 keV, $B = 54$~MG. The $\otimes$ 
denotes an overcorrected atmospheric feature.}
\end{figure}

\sect{The faint-phase cyclotron spectrum}

The spectrum during the faint phase consists of photospheric radiation 
from the stars, line and continuum radiation from the accretion stream, 
and cyclotron radiation from the secondary accretion region. In order 
to isolate the cyclotron component and to compare it with cyclotron 
models, we subtracted a smooth contribution from the faint-phase 
spectra in Fig.~\ref{mean_spec}
 which approximates the non-cyclotron components. 
It was calculated by spline fits to continuum data points selected 
at the cyclotron minima. The residuals shown in Fig.~\ref{cyc_fai} were than 
compared with homogeneous cyclotron models (Chanmugam \& Dulk 1981; 
Schwope et al. 1990). A nice fit was obtained assuming kT = 5 keV, 
$B_2 = 54$~MG, $\theta = 70^\circ$, and $\log \Lambda = 3.2$. The 
corresponding model curve is also shown in Fig.~\ref{cyc_fai}. 
Since the IDS and 
CCD faint-phase spectra differ just by a wavelength-independent factor, 
the spectral parameters (kT, $B,\ \theta,\ \Lambda$) apply to both 
observations. WFB arrived at nearly identical values (kT = 5 keV, 
$B = 54.6$~MG, $\theta = 75^\circ$ and $\log \Lambda = 3.04$) for their 
data which indicates that we were observing the same region on the white 
dwarf. 

For the above spectral parameters, the cyclotron luminosity of the 
second pole obtained from the CCD-data is $L_{\rm cyc,2} \simeq 3 \times 
10^{29} d_{100}^2\,$erg~s$^{-1}$, the cyclotron emitting area 
$A_{\rm cyc,2} \simeq 2 \times 10^{14} d_{100}^2\,$cm$^2$ and the total 
and specific mass accretion rates $\dot M_{\rm 1989} \simeq 3 \times 
10^{12} R_{8.9}\,M_{0.6}^{-1}\,d_{100}^2\,$g~s$^{-1}$ and $\dot m_2 \simeq 
0.015\ d_{100}^2\,$g~cm$^{-2}$s$^{-1}$, respectively ($d_{100}$: distance 
in units of 100 pc; $R_{8.9}$: white dwarf radius in units of $10^{8.9}\,$ 
cm; $M_{0.6}$: white dwarf mass in units of $0.6 M_\odot$). For the 
IDS-data, the values of $L, A$ and $\dot M$ have to be divided 
by a factor of 2. The mass accretion rate has been estimated by assuming 
$L_{\rm acc} \simeq L_{\rm cyc}$, as is suggested by the non-detection 
of soft X-rays during the faint phase. The spectral parameters, luminosity 
and mass accretion rate of the second pole in VV Pup are very similar to 
those of UZ For in a low state of accretion (Schwope et al. 1990). Hence, 
the secondary accretion region is more likely heated via particle 
bombardement than by a hydrodynamic shock (Woelk \& Beuermann 1992). 

\begin{figure}[t]
\begin{center}
\begin{minipage}{88mm}
\psfig{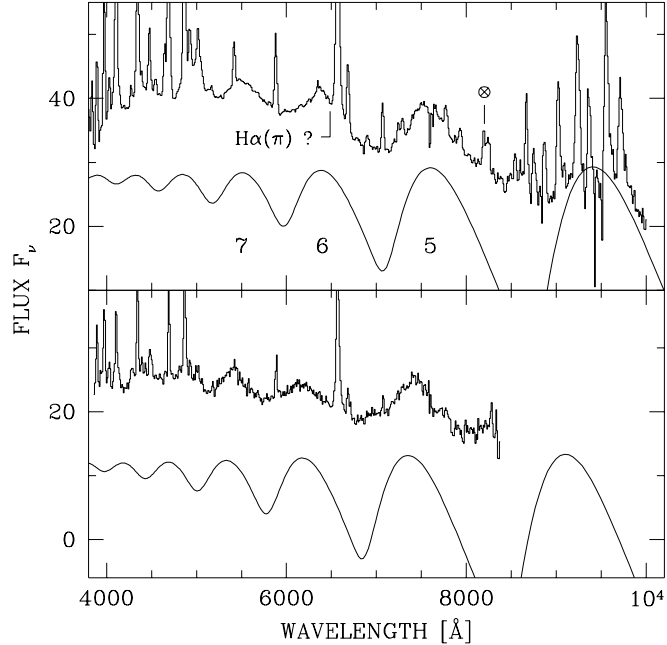}
\end{minipage}
\end{center}
\caption{\label{cyc_bri}
Cyclotron spectra of the main accretion region obtained in 
Jan. 1989 (CCD-data, upper panel) and Febr. 1984 (IDS-data, lower panel). 
Below the data normalized cyclotron absorption coefficients for a 
10 keV-plasma seen at $\Theta = 75^\circ$ are shown (upper panel: 
$B = 30.5$~MG, lower panel: $B = 31.5$~MG). The harmonics used for 
determination of the field strength are indicated by integers. The 
symbol $\otimes$ denotes an overcorrected atmospheric features.}
\end{figure}

\sect{The bright-phase cyclotron spectrum}
In order to extract the spectral flux from the main pole all other 
radiation components have to be subtracted, including cyclotron 
radiation from the always visible secondary region. Since the secondary 
cyclotron lines obviously move with phase, 
it is not allowed simply to subtract 
a mean faint-phase spectrum. Instead we subtracted from the bright-phase 
spectra shown in Fig.~\ref{mean_spec} the smooth continuum already used for 
the faint-phase spectra and a cyclotron model spectrum calculated using 
the same spectral parameters as for the faint-phase spectrum 
(Fig.~\ref{cyc_fai}), 
except a polar angle of $\theta = 85^\circ$ instead of $70^\circ$. The 
small change in $\theta$ accounts for the observed motion of the lines. 
The residuals after subtraction are shown in Fig.~\ref{cyc_bri}.

Cyclotron lines are clearly visible in both sets of data and they are 
also clearly shifted with respect to each other. These lines occur in a 
spectral region where cyclotron emission is usually thought to be 
optically thick (see e.g. Wickramasinghe, 1988, for a comprehensive 
review on different types of cyclotron models). The situation is 
similar to V834 Cen (Schwope \& Beuermann 1990) where we took the 
simple view that the spectrum consists of an optically thick cyclotron 
continuum with emission from some optically thin region superimposed. 
We take the same view here and assume that the optically thin region 
is isothermal. Its magnetic field strength may then be determined 
by comparing the observed spectra with the cyclotron absorption coefficient 
$\kappa_{\rm cyc}$ for given temperature and polar angle.

For practical purposes, we used normalized $\kappa$'s, obtained by 
division by a smooth curve connecting all harmonic maxima. Examples 
are shown also in Fig.~\ref{cyc_bri}. We fitted only the $5^{\rm th}$ 
to $7^{\rm th}$ 
harmonics because the $4^{\rm th}$ and the $8^{\rm th}$ to $10^{\rm th}$ 
harmonics which are also covered by our spectral range lie below strong 
emission lines. Assuming kT = 10 keV and $\theta = 75^\circ$, we 
obtain field strengths $B_{\rm 1,CCD} = 30.5$~MG and $B_{\rm 1,IDS} =
31.5$~MG. In both cases, we fixed the polar angle at $75^\circ$ in 
accordance with several polarimetric studies of VV Pup (Brainerd \& Lamb 
1985; Cropper \& Warner 1986; Piirola et al. 1990). All authors 
consistently derived $i \simeq 75^\circ$ and $\eta_1 \simeq 150^\circ$, 
so that for the center of the bright phase $\theta = \eta - i \simeq 
75^\circ$. If one assumes a temperature of 5 keV, the field strengths 
given above have to be lowered by $\sim 1$~MG. Lower temperatures are 
excluded because they cannot reproduce the phase-dependent motion of 
cyclotron lines reported here. Various authors have determined $B_1$ and 
their results are compiled in Table 1 together with references to the 
data used. The large spread in the tabulated data, $B_1 = 27.4 - 
31.8$~MG is based both, on a variety of model assumptions, and on
real changes in the physical conditions of the accretion 
region. Our observation of a wavelength shift of the cyclotron lines between 
1984 and 1989 finds an explanation in a change of either of the parameters 
$B_1, T$, and $\theta$ or in a combination of them and we comment on them 
subsequently. 

\begin{table}[t]
\begin{center}
\begin{minipage}{90mm}
\caption[]{Field strength $B_1$ of the primary pole}
\begin{tabular}{cccll}
\hline\noalign{\smallskip}
$B_1$ & kT & $\Theta$ & Data & Reference \\
(MG) & (keV) &        & & \\
\hline
31.8 & 10 & $90^\circ$ & SLB, WV & WM82 \\
31.5 & 8.7 & $85^\circ - 90^\circ$ & SLB, WV & BC85 \\
27.4--30.8 & 0.2--2.5 & $85^\circ$ & SLB & CO88\\
30.5 & 10 & $85^\circ$ & WFB & WFB \\
30.5 & 10 & $75^\circ$ & CCD & this work \\
29.5 & 5 & $75^\circ$ & CCD & this work \\
31.5 & 10 & $75^\circ$ & IDS & this work \\
30.5.5 & 5 & $75^\circ$ & IDS & this work \\
\noalign{\smallskip}
\hline
\end{tabular}
VW: Visvanathan \& Wickramasinghe (1979); WV: Wickramasinghe \& Visvanathan
(1980); WM82: Wickramasinghe \& Meggitt (1982); BC85: Barrett \& Chanmugam 
(1985); CO88: Canalle \& Opher (1988)
\end{minipage}
\end{center}
\end{table}

A pure change in $\theta$ would result in different lengths of the bright 
phases of the IDS- and the CCD-data which are not observed here, but which are
present in a statistical sense in the combined data from the literature
(see below).

A change in $B_1$ was first proposed by WFB (Table 1, 1{\em st} and 4{\em 
th} row) 
who interpreted this in terms of a migration of the centroid of the 
emission region in response to a change in accretion rate. Taking this 
view implies that 
the accretion stream is able to penetrate deeper into 
the magnetosphere when the accretion rate increases. The lower field 
seen in the CCD data corresponds then to a motion in stellar latitude 
of $\Delta\vartheta = 10^\circ - 15^\circ$ towards the magnetic equator. 
We note that the shift in $B_1$ (if real), is not 
accompanied by a corresponding shift in $B_2$. This is suggestive 
of a geometry where both accretion regions 
are not located at the footpoints of 
the same closed field line(s). The ways of both accretion streams 
must be to a large extent independent of each other while the accretion 
rates are not. The levels of cyclotron activity at both poles 
increase simultaneously.

\begin{figure}
\begin{center}
\begin{minipage}[]{100mm}
\psfig{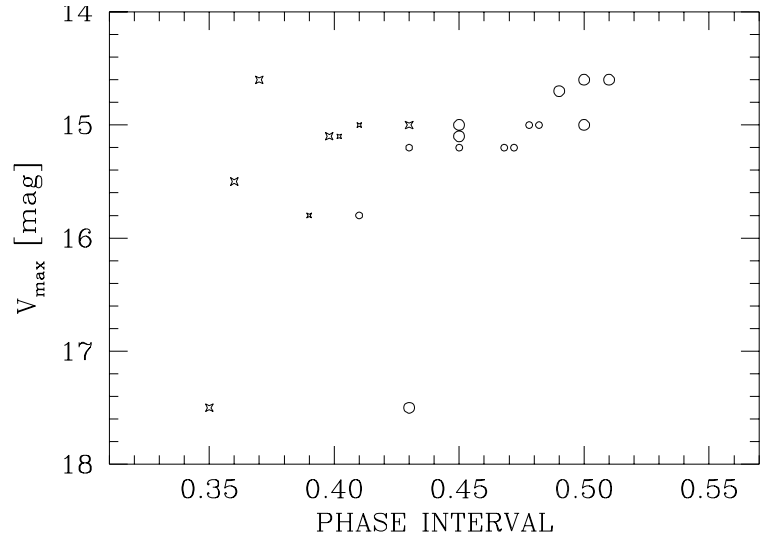}
\end{minipage}
\end{center}
\caption{\label{length}
Length of bright phase $\Delta\phi_{\rm B}$ (o) and phase 
interval of positive circular polarization $\Delta\phi_{\rm CP}$ (x) 
as a function of maximum brightness. Large symbols denote observations 
with measured magnitudes, small symbols has been used to denote data 
where the brightness has been estimated.}
\end{figure}

If the whole accretion spot is displaced with varying accretion rate by 
$\Delta\vartheta = 10^\circ - 15^\circ$, one would expect that this 
behaviour is also reflected in the length of the bright phase 
$\Delta\phi_{\rm B}$ or in the length of the phase interval of positive 
circular polarization $\Delta\phi_{\rm CP}$. The expected change is 
$\sim 0.1$ phase units. In Table~2, we have compiled all available data 
for $\Delta\phi_{\rm B}$ and $\Delta\phi_{\rm CP}$ found in the 
literature and, in Fig.~\ref{length}, 
we give a graphical representation of these 
data. We use there the maximum orbital brightness as estimator of 
the mass accretion rate. 
The value of $\Delta\phi_{\rm CP}$ seems to be without trend, 
probably due to varying contributions from the second pole with opposite 
field polarity. The length of the bright phase indeed displays a 
variation of about $\Delta\phi_{\rm B} \simeq 0.1$ phase units with the 
more extended bright phase preferentially occuring at higher system 
brightness. But the relation is not unique in the sense that it indicates a
lateral and/or radial change of the size of the accretion spot as a further 
complication. This applies particularly to the IDS- and CCD-observations 
presented here, which have equal lengths $\Delta\phi_{\rm B}$ within 
the errors. A migration of the accretion region with the corresponding 
shift in $B_1$ alone cannot, therefore, explain the whole spectroscopic 
and photometric information.

As a third possibility, the cyclotron lines might experience a  
redshift caused by a 
temperature increase in response to an enhanced mass transfer rate. 
The line positions of the CCD- and the IDS-spectrum may be reproduced 
assuming equal field strength, $B_1 = 30.5$~MG, but different plasma
temperatur, $kT_{\rm CCD} \simeq
10$~keV and $kT_{\rm IDS} \simeq 5$~keV, respectively (see Table~1). 
Hence, combined $T$- and $B$-variations along with variations in the extent 
of the accretion region  appear as the likely mechanism to explain the 
shift of the cyclotron line system.

The field configuration of the white dwarf in VV Pup remains uncertain at
present. For a dipolar field, the observed ratio $B_1/B_2$ would 
require a decentering along the dipole axis of 0.1  white dwarf radii. 
The reasons which cast doubt on the dipolar assumption, on the other hand, 
are explained in section 3.2. It would, therefore, be important to 
obtain high-quality spectra during a low state which would allow a 
phase-dependent study of the photospheric field to be performed.

The bright-phase CCD-spectrum shown in Fig.~\ref{cyc_bri} 
displays an intensity 
minimum blueside of the strong $H\alpha$-emission line. This can also
be recognized in the grey-scale representation of Fig.~\ref{grey_spec}.
It is labeled 
$H\alpha(\pi)$ and may be interpreted as absorption of nonphotospheric 
origin due to the presence of cool matter in the vicinity of the hot 
emission region (as first seen in V834 Cen, Wickramasinghe et al. 1987). 
Assuming a field strength in the absorption region of $\ga 30$~MG, the 
positions of the corresponding $\sigma$-components blend with the 
cyclotron minima between the $5^{\rm th}, 6^{\rm th}$, and $7^{\rm th}$ 
harmonics and are probably not detected for that reason. 

Finally, we comment on the optically thick part of the cyclotron 
spectrum. Homogeneous cyclotron models are not suitable to match 
spectra which are as flat as those observed and which, in addition, 
display optically thin cyclotron lines in the bona-fide optically 
thick long-wavelength range. However, the simple model spectra may 
serve as a rough guide in order to obtain average values for the plasma 
parameters of the structured emission region. Assuming kT$_{\rm cyc} = 
10 - 20$~keV, the spectral maximum around 4000 \AA\ is reproduced for 
optical depth parameters $\log \Lambda = 6.3 - 5.3$. The integrated 
cyclotron flux obtained from the IDS spectrum (Fig.~\ref{cyc_bri}) 
multiplied by 
a factor of 1.5 to account for those parts lying outside our spectral 
window is $F_{\rm cyc} \sim 1.5 \times 10^{-12}\,$erg~cm$^{-2}$s$^{-1}$. 
>From the simultaneous EXOSAT-observations (count rate ratio Lexan 3000/AlP 
= 11 $\pm$ 3, no ME-detection, Osborne et al. 1984), blackbody and 
bremsstrahlung fluxes may be estimated as $F_{\rm bb} = (1-10^3) \times 
10^{-10}\,$erg~cm$^{-2}{\rm s}^{-1}$ and $F_{\rm br} < 1.5 \times 
10^{-12}\,$erg~cm$^{-2}{\rm s}^{-1}$, respectively. Thus, there is a 
handsome soft X-ray excess while the bremsstrahlung contribution does 
not exceed the cyclotron flux. The soft X-ray excess is probably due to 
the infall of dense clumps or filaments of matter
(Frank et al. 1988) and is restricted to 
the primary pole. Hence, two substantially differing modes of energy 
release and cooling are acting in both accretion regions.

\begin{table}[t]
\begin{center}
\begin{minipage}[b]{99mm}
\caption{Length of bright phase as a function of brightness}
\begin{tabular}{lllccl}
\hline
Date & $V_{\rm max}$ & $V_{\rm min}$ & $\Delta\phi_{\rm B}$ & $\Delta\phi_{\rm CP} $ & Reference \\
(Y/M/D)  & \multicolumn{2}{c}{(mag)} &&& \\
\hline
\noalign{\smallskip}
unknown     &15.5  &      &---        &0.36 &BL85 \\
70/11/05    &15.2: &16.5: &0.47       &---  &WN72 \\
71/01/20    &15.2: &16.5: &0.43, 0.45 &---  &WN72 \\
71/01/25    &15.2: &16.5: &0.47       &---  &WN72 \\
77/11/06    &17.5  &18    &0.43       &---  &B78 \\
77/12/06    &17.5  &      &---        &0.35 &Lea78 \\
78/03/04    &15:   &      &0.48       &0.41 &LS79 \\
79/12/15    &14.6  &15.9  &0.51       &---  &Sea83 \\
84/02/1-4   &15    &16.8  &0.45       &0.43 &CW86 \\
84/02/05    &15    &16.8  &0.50       &---  &IDS-data \\
85/02/13-14 &15.8: &      &0.41       &0.39 &CW86 \\
86/01/06    &15.0: &16.5: &---        &0.40 &B88 \\
86/01/10-19 &15.1  &16.5  &0.45       &0.40 &Pea90 \\
87/03/21    &15:   &      &0.48       &---  &L88 \\
88/02/18-26 &14.6  &15.7  &0.50       &0.37 &Pea90 \\
89/01/18    &14.7  &16.0  &0.49       &---  &CCD-data \\
\noalign{\smallskip}
\hline
\end{tabular}
\noindent 
BL85: Brainerd \& Lamb 1985; WN72: Warner \& Nather 1972; B78: Bailey 1978;
Lea78: Liebert at al.~1978; LS79: Liebert \& Stockman 1979; Sea83: Szkody
et al.~1983; CW86: Cropper \& Warner 1986; B88: Bailey 1988; Pea90: Piirola 
et al.~1990; L88: Larsson 1988\\
\noindent
Magnitudes without colon indicate measured values. Magnitudes with colon are
estimated from statements about high or low accretion states or relative 
brightness
comparing different observations. 
\end{minipage}
\end{center}
\end{table}

\kap{Conclusions}
We have studied the cyclotron spectra from the two accretion regions on the 
white dwarf in VV Pup. The main results are based on the determination of 
field strengths in both emission regions and on the phase-dependent 
motion of cyclotron lines. They may be summarized as follows:
\begin{itemize}
\item[$\bullet$]The accretion regions most probably lie not at the 
footpoints of the same closed field line or bundle of field lines.
\item[$\bullet$]The field configuration
deviates from a centered dipole.
\item[$\bullet$]The cyclotron lines from the main (near) accreting pole 
display a shift between 1984 and 1989 while the lines from the second 
(far) pole do not. The shift is caused either by a deeper penetration 
of the accretion stream into the magnetosphere and corresponding decrease 
of field strength or by a temperature increase with increasing mass 
accretion rate.
\item[$\bullet$]The two poles differ in their emission properties and, 
hence, in the underlying heating and cooling processes.
\end{itemize}

{\it Acknowledgements.}\
We thank an anonymous referee for helpful comments.
This work was supported by the BMFT under grants 50 OR 9101 5 and 
50 OR 9403 5.

\refer
\aba

\rf{Bailey J., 1978, Mon.~Not.~R.~Astron.~Soc., 185, 73P}
\rf{Bailey J., 1988, 
     In: Polarized Radiation of Circumstellar Origin, Coyne G.V. et al. (eds.) 
  	Univ. Ariz. Press, \\Tucson, p.~105}
\rf{Barrett P.E., Chanmugam, G., 1985, Astrophys.~J., 298, 743}
\rf{Brainerd J.J., Lamb, D.Q., 1985, 
     In: Lamb D.Q., Patterson, J. (eds.) Proc. 7$^{th}$ North American   
     Workshop on CV's and LMXBs, Reidel, Dordrecht, p.~247}
\rf{Burwitz V., Reinsch K., Schwope A.D., Beuermann K., Thomas H.-C.,
	Greiner J., 1996, Astron.~Astrophys., 305, 507}
\rf{Canalle J.B.G., Opher R., 1988, Astron.~Astrophys., 189, 325}
\rf{Chanmugam G., Dulk G.S., 1981, Astrophys.~J., 244, 569}
\rf{Cropper M., 1988, Mon.~Not.~R.~Astron.~Soc., 231, 597}
\rf{Cropper M., Warner B., 1986, Mon.~Not.~R.~Astron.~Soc., 220, 633}
\rf{Frank J., King A.R., Lasota J.-P., 1988, Astron.~Astrophys., 193, 113}
\rf{Larsson S., 1988, Adv. Space Res., 8(2), 305}
\rf{Liebert J., Stockman H.S., 1979, Astrophys.~J., 229, 652}
\rf{Liebert J., Stockman H.S., Angel J.R.P., Woolf N.J., Hege K. 
     Morgan B., 1978, Astrophys.~J., 225, 201}
\rf{Osborne J., Maraschi L., Beuermann K., et al., 1984, 
     In: Oda M., Giacconi R. (eds.), X-ray Astronomy '84, \\Bologna, p.~59}
\rf{Piirola V., Coyne G.V., Reiz A., 1990, Astron.~Astrophys., 235, 245}
\rf{Schwope A.D., 1990, Reviews in Modern Astronomy, 3, 44}
\rf{Schwope A.D., Beuermann K., Jordan S., Thomas H.-C., 
	1993, Astron.~Astrophys., 278, 487}
\rf{Schwope A.D., Thomas H.-C., Beuermann K., Burwitz V.,
	Jordan S., Haefner R., 1995, Astron.~Astrophys., 293, 764}
\rf{Stockman H.S., Liebert J., Bond H.E., 1979, 
	In: van Horn H.H. \& Weidemann V. (eds), Proc. IAU Coll.~53, 
	Univ. Rochester, p. 334 (SLB)}
\rf{Szkody P., Bailey J.A., Hough J.H., 1983, 
    Mon.~Not.~R.~Astron.~Soc., 203, 749}
\rf{Tapia S., 1977, IAU Circ. No. 3045}
\rf{Visvanathan N., Wickramasinghe D.T., 1979, Nature, 281, 47}
\rf{Walker M.F., 1965, Mitt. Sternw. Budapest 57}
\rf{Warner B., Nather R.E., 1972, Mon.~Not.~R.~Astron.~Soc., 156, 305}
\rf{Wickramasinghe D.T., 1988, 
     In: Polarized Radiation of Circumstellar Origin, Coyne G.V. et al. (eds.),
	Univ. Ariz. Press, Tucson, p.~199}
\rf{Wickramasinghe D.T., Meggitt S.M.A., 1982, 
    Mon.~Not.~R.~Astron.~Soc., 198, 975}
\rf{Wickramasinghe D.T., Visvanathan N., 1980, 
    Mon.~Not.~R.~Astron.~Soc., 191, 583}
\rf{Wickramasinghe D.T., Ferrario L., Bailey J., 1989, Astrophys.~J., 342, L37}
\rf{Wickramasinghe D.T., Tuohy I.R., Visvanathan N., 1987, 
    Astrophys.~J., 318, 326}
\rf{Woelk U., Beuermann K., 1992, Astron.~Astrophys.~256, 498}
\rf{}

\abe

\address
A.D.~Schwope\\
Astrophysikalisches Institut Potsdam, An der Sternwarte 16,
D--14482 Potsdam\\
Germany\\
E--mail: ASchwope@aip.de \\[2mm]
Klaus Beuermann\\
Universit\"{a}tssternwarte G\"{o}ttingen\\
D--37083 G\"{o}ttingen, Germany and\\
MPI f\"ur extraterrestrische Physik\\
D--85740 Garching b. M\"unchen\\
Germany
END

\end{document}